\documentclass[conference,10pt]{IEEEtran}
\IEEEoverridecommandlockouts
\usepackage{cite}
\usepackage{amsmath,amssymb,amsfonts}
\usepackage{algorithm}
\usepackage{algpseudocode}
\usepackage{graphicx}
\usepackage{textcomp}
\usepackage{xcolor,soul}
\usepackage{bm}
\usepackage{subfigure}
\def\BibTeX{{\rm B\kern-.05em{\sc i\kern-.025em b}\kern-.08em
    T\kern-.1667em\lower.7ex\hbox{E}\kern-.125emX}}

\begin{document}

\title{Neuromorphic In-Context Learning for Energy-Efficient MIMO Symbol Detection}

\author{\IEEEauthorblockN{Zihang Song,  Osvaldo Simeone, and Bipin Rajendran}
\IEEEauthorblockA{Centre for Intelligent Information Processing Systems, Department of Engineering\\ King's College London, London WC2R 2LS, U.K.}
\thanks{This work is supported in part by the European Union’s Horizon Europe project CENTRIC (101096379), the EPSRC project (EP/X011852/1) and by Open Fellowships of the EPSRC (EP/W024101/1 and EP/X011356/1).}
\thanks{Corresponding author: Bipin Rajendran (bipin.rajendran@kcl.ac.uk).}
}

\maketitle

\begin{abstract}
In-context learning (ICL), a property demonstrated by transformer-based sequence models, refers to the automatic inference of an input-output mapping based on examples of the mapping provided as context. ICL requires no explicit learning, i.e., no explicit updates of model weights, directly mapping context and new input to the new output.  Prior work has proved the usefulness of ICL for detection in MIMO channels. In this setting, the context is given by pilot symbols, and ICL automatically adapts a detector, or equalizer, to apply to newly received signals. However, the implementation tested in prior art was based on conventional artificial neural networks (ANNs), which may prove too energy-demanding to be run on mobile devices. 
This paper evaluates a neuromorphic implementation of the transformer for ICL-based MIMO detection. This approach replaces ANNs with spiking neural networks (SNNs), and implements the attention mechanism via stochastic computing, requiring no multiplications, but only logical AND operations and counting. When using conventional digital CMOS hardware, the proposed implementation is shown to preserve accuracy, with a reduction in power consumption ranging from $5.4\times$ to $26.8\times$, depending on the model sizes, as compared to ANN-based implementations.

%approach utilizing spiking neural networks (SNNs) to reduce power consumption while enabling end-to-end MIMO symbol detection. Our method leverages the low-power characteristics of SNNs in conjunction with the dynamic adaptability provided by the ICL capability of transformer-based sequence models, enabling them to dynamically adjust to unknown channel conditions through the contextual analysis of received pilot data. T This work advances the development of power-efficient adaptive AI receivers for future 6G networks, particularly in scenarios characterized by complex channel conditions and stringent power constraints.
\end{abstract}

\begin{IEEEkeywords}
MIMO symbol detection, Transformers, In-context learning, Neuromorphic computing, Energy efficiency
\end{IEEEkeywords}

\section{Introduction}

The integration of machine learning into wireless communication links is expected to significantly boost the capabilities of intelligent receivers, representing a notable step towards 6G systems \cite{you2021towards}. A key challenge in the design of wireless receivers is ensuring efficient adaptation to changing channel conditions. \emph{In-context learning} (ICL), a property demonstrated by transformer-based sequence models, refers to the automatic inference of an input-output mapping based on examples of the mapping provided as context \cite{garg2022can}. ICL requires no explicit learning or optimization, i.e., no explicit updates of model weights, directly mapping context and new inputs to new outputs.  Prior work has proved the usefulness of ICL for detection in MIMO channels \cite{rajagopalan2023transformers,zecchin2023context}. In this setting, the context is given by pilot symbols, and ICL automatically adapts a detector, or equalizer, to apply to newly received signals.

 Despite the performance benefits demonstrated in prior art \cite{rajagopalan2023transformers,zecchin2023context}, the execution of a transformer model requires substantial computational resources, potentially increasing the power consumption of a receiver to levels that are impractical for deployment on edge devices. Consequently, optimizing the power efficiency of transformer models is a key step in making ICL a broadly applicable technique for next-generation receivers.

\begin{figure}[t]
    \centering
    \includegraphics[width=8cm]{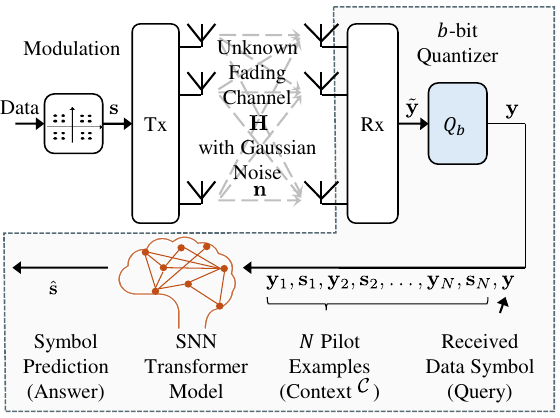}
    \vspace{-3mm}
    \caption{Illustration of in-context learning  (ICL) for MIMO symbol detection.}
    \label{fig1}
    \vspace{-5mm}
\end{figure}

Inspired by the neural architecture of human brains, which operate on as little as 20 Watts --, \textit{neuromorphic computing} offers a compelling solution to these challenges \cite{mead1990neuromorphic}. Unlike the traditional artificial neural network (ANN)  paradigm, neuromorphic computing relies on \textit{spiking neural networks} (SNNs), which adopt a time-encoded, event-driven approach that closely resembles biological neural processing \cite{tavanaei2019deep}. SNNs have been proven to offer energy savings for many workloads due to their sparse, multiplication-free, calculations \cite{Davies}.

Integrating SNNs into wireless communication systems offers a promising solution to the critical issue of energy consumption, and has been previously investigated for semantic communications \cite{skatchkovsky2021spiking,chen2022neuromorphic}, integrated sensing and communications \cite{chen2023neuromorphic}, and satellite communications \cite{ortiz2024energy}.

As illustrated in Fig.~\ref{fig1}, this paper evaluates a neuromorphic implementation of the transformer for ICL-based MIMO detection. This approach replaces ANNs with SNNs and implements the attention mechanism via stochastic computing \cite{gaines1969stochastic}, requiring no multiplications, but only logical AND operations and counting \cite{song2024stochastic}.  The proposed implementation is tested for MIMO detection in a link with a quantized receiver frontend.  When using conventional digital CMOS hardware,
the proposed implementation is shown to preserve accuracy, with
a reduction in power consumption ranging from 5.4$\times$ to 26.8$\times$, depending on the model sizes, as compared to ANN-based implementations.

% In this study, we introduce an innovative design that applies a neuromorphic paradigm to transformer-based models to address the challenge of end-to-end symbol detection in MIMO systems. Our findings reveal that integrating SNNs with the transformer architecture not only achieves high performance but also reduces computational complexity and power consumption. This integration enables the system to leverage limited contextual information for enhanced performance in novel scenarios, without necessitating extensive retraining. Moreover, it achieves high demodulation accuracy with significantly reduced energy consumption. The key contributions of this paper are outlined as follows:
% \begin{enumerate}
%     \item We apply ICL to end-to-end symbol detection in unknown MIMO channels, leveraging low-precision processing at the receiver. Unlike traditional models that rely on fine-tuning with pilot sequences, our ICL methodology interprets transmitted symbols directly from received signals by using pilot vector-symbol pairs for contextual guidance. This method facilitates streamlined adaptation across varying channel scenarios without necessitating alterations to model parameters.

%     \item Distinct from traditional ANN approaches, we develop a neuromorphic SNN transformer model to efficiently conduct task 1) in a low-precision and low-computational complexity setup, thereby achieving significant energy efficiency. 
% \end{enumerate}

\section{Problem Formulation}\label{se_model}

As in \cite{zecchin2023context}, we consider a standard \(N_t \times N_r\) MIMO link with $N_t$ transmit antennas, $N_r$ receive antennas, and a quantized front-end receiver. The transmitted data vector \(\mathbf{s}\) consists of $N_t$ entries selected independently and uniformly from a constellation $\mathcal{S}$ with  \(K\) symbols. The average transmitting power is given by \(\mathbb{E}[\mathbf{\|s\|}^2]=N_t\). The unknown fading channel is described by an $N_r\times N_t$  matrix \(\mathbf{H}\), yielding the $N_r\times 1$ pre-quantization received signal vector
\begin{equation}\label{eq:rx0}
\tilde{\mathbf{y}} = \mathbf{H}\mathbf{s}+ \mathbf{n},
\end{equation} with additive white complex Gaussian noise $\mathbf{n}$ having zero mean and unknown variance \(\sigma^2\). Upon $b$-bit quantization, the received  signal \(\mathbf{y}\) available at the detector is written as 
\begin{equation}\label{eq:rx}
\mathbf{y} = Q_b(\tilde{\mathbf{y}}),
\end{equation}
where the function $Q_b(\cdot)$ clips the received signal within a given range $[l_{\text{min}},l_{\text{max}}]$ and applies uniform quantization with a resolution of $b$ bits separately to the real and imaginary parts of each entry.

The objective of symbol detection is to estimate the transmitted data vector \(\mathbf{s}\) from the corresponding received quantized signal vector \(\mathbf{y}\). The detector does not know the channel and noise level $(\mathbf{H},\sigma^2)$. Rather, it has access only to a context $\mathcal{C}$, which comprises of $N$ independent and identically distributed (i.i.d.) pilot receive-transmit pairs
\begin{equation}\label{eq:context}
    \mathcal{C}=\{(\mathbf{s}_i,\mathbf{y}_i)\}^N_{i=1}
\end{equation}obtained from the channel model  (\ref{eq:rx0})-(\ref{eq:rx}).

\section{In-Context Learning for MIMO Symbol Detection}

Recently, ICL was proposed as a mechanism to  estimate the symbol vector $\mathbf{s}$ for a new received signal $\mathbf{y}$ based on the context $\mathcal{C}$ in (\ref{eq:context}) \cite{zecchin2023context}. Note that both the data input-output pair $(\mathbf{s},\mathbf{y})$ and the pilots in the context  $\mathcal{C}$ are assumed to be generated i.i.d. from the model (\ref{eq:rx0})-(\ref{eq:rx}). As illustrated in Fig. 1, in this approach, a transformer model is fed the context $\mathcal{C}$ and the received signal $\mathbf{y}$ of interest and is trained to output an estimate $\hat{\mathbf{s}}$ of the transmitted signal vector $\mathbf{s}$. 

Training is done by collecting a data set consisting of pairs of context $\mathcal{C}$ and transmitting data-received signal pair $(\mathbf{s},\mathbf{y})$. Each such data point is obtained for a given realization of the channel $\mathbf{H}$ and channel noise $\sigma^2$. Specifically, for each realization $(\mathbf{H},\sigma^2)$, referred to as a task, $N_{\text{Example}}$ context-data pairs are generated and included in the data set (see Sec. \ref{se_model}). 

As shown in \cite{zecchin2023context}, and further discussed in Sec. \ref{se_eval}, ICL can approach the performance of an ideal minimum mean squared error (MMSE) estimator that knows the statistics of channel $\mathbf{H}$, as well as the channel noise power $\sigma^2$. However, the implementation in \cite{zecchin2023context} uses conventional ANN models, resulting in potentially prohibitive computing energy consumption levels. In this paper, we focus on evaluating the performance of a more efficient implementation of the transformer based on neuromorphic computing, i.e., on SNNs \cite{song2024stochastic}. 

\section{Preliminaries}

This section introduces important background information for the definition of the considered SNN-based transformer.

\subsection{Bernoulli Encoding and Stochastic Computing}
Unlike ANNs, SNNs require the input data to be encoded in binary temporal sequences. In this work, Bernoulli encoding is adopted to convert a normalized real value \(\overline{x} \in [0,1]\) into a discrete binary sequence of length \(T\). This is done by assigning each encoded bit the value of `1' independently with probability \(\overline{x}\), and the value `0' with probability \(1 - \overline{x}\), which is denoted as
\begin{equation}\label{eq_bernoulli}
    x^t \sim \text{Bern}(\overline{x}),\; t = 1, 2, \ldots, T.
\end{equation}
Bernoulli encoding enables the execution of stochastic computing \cite{gaines1969stochastic}. 

In stochastic computing, the multiplication between two numbers $\overline{x}\in[0,1]$ and $\overline{y}\in[0,1]$ can be efficiently approximated by using the corresponding Bernoulli encoded sequences as
\begin{equation}
    z^t = x^t \land y^t,\; t = 1, 2, \ldots, T,
\end{equation}
thus utilizing a simple logic AND gate $\land$. In fact, the probability of obtaining an output equal to 1 is equal to the desired product, i.e.,
$P(z^t = 1) = \overline{x} \times \overline{y}$.

\subsection{Leaky Integrate-and-Fire Neurons}
In SNNs, neurons maintain an internal analog membrane potential and emit a spike when the potential crosses a threshold. A leaky integrate-and-fire (LIF) spiking neuron integrates incoming signals over discrete time steps. The membrane potential \( V^t \) of an LIF neuron at time step \( t \) evolves as
\begin{equation}\label{eq:V}
\begin{aligned}
V^t &= \beta V^{t-1} + I^t,\\
\end{aligned}
\end{equation}
where \( \beta\in(0,1] \) represents the leak factor determining the rate of decay of the membrane potential over time, and \( I^t \) is the input at time step \( t \). When \( V^t \) reaches or exceeds a threshold voltage \( V_{\text{thresh}} \), the neuron fires a spike, i.e., $O^t=1$, resetting the potential $V^t$ to $0$. Otherwise, the neuron remains silent, producing $O^t=0$, i.e., 
\begin{equation}\label{eq:th}
\text{if } V^t \geq V_{\text{thresh}} \text{ then }
\begin{cases}
O^t = 1,\\
V^t = 0,
\end{cases}
\text{ else } O^t = 0 \text{ and \eqref{eq:V}}.
\end{equation}

A layer of LIF neurons takes a binary vector $\mathbf{x}_{\text{in}}^t$, applies a linear transformation via a trainable weight matrix $\mathbf{W}$ and then applies the dynamics \eqref{eq:th} separately to each output. Accordingly, the output of a LIF layer at time $t$ is denoted as \begin{equation}\label{eq:LIFlayer}
   \mathbf{x}_{\text{out}}^{t} = \text{LIF}^{t}(\mathbf{W} \mathbf{x}_{\text{in}}^t), 
\end{equation}
The notation $\text{LIF}^{t}(\cdot)$ in (\ref{eq:LIFlayer}) implicitly accounts for the dependence of the output $\mathbf{x}_{\text{out}}^{t}$ also on the past inputs through the membrane potentials \eqref{eq:th} of the spiking neurons.  Importantly, since the input signal  \(\mathbf{x}_{\text{in}}^t\) is binary, the matrix-vector multiplication $\mathbf{W} \mathbf{x}_{\text{in}}^t$ can be efficiently evaluated without the need to implement multiplications. This is one of the key reasons for the energy efficiency of SNNs as compared to ANNs.

% \subsection{SNN Training}
% At the final output stage of the SNN, the spike trains are averaged across all time steps, yielding a real-valued output. Subsequently, the loss is computed analogously to that in traditional ANNs. Stochastic gradient descent (SGD) is employed to optimize the parameters for each mini-batch. Despite the LIF neuron lacking an explicit gradient function, a \textit{sigmoid} function is utilized as a surrogate gradient, simplifying the training process.

\section{Spiking transformer for MIMO Symbol Detection}
\begin{figure}[t]
    \centering
    \includegraphics[width=8.7cm]{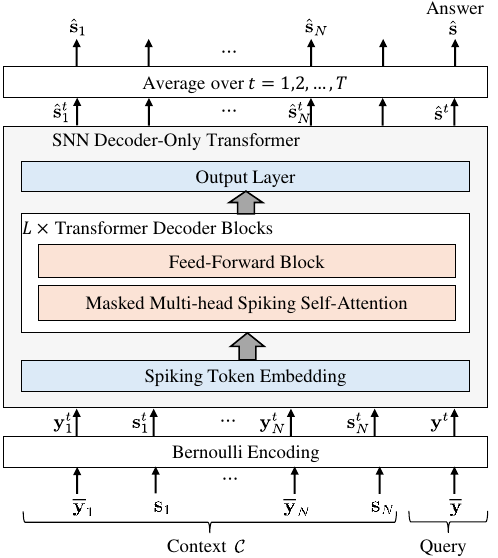}
    \caption{Block diagram of the proposed spiking transformer for ICL-based MIMO symbol detection.}
    \label{model}
    \vspace{-5mm}
\end{figure}

In this section, we detail the proposed implementation of MIMO symbol detection using a decoder-only spiking transformer. The proposed scheme, illustrated in Fig.~\ref{model}, replaces the conventional ANN architecture adopted in \cite{zecchin2023context}, with a fully spiking neuron-based architecture that adopts a stochastic self-attention module \cite{song2024stochastic}.

\subsection{Normalization and Spike Encoding}

The transformer model in Fig. 2 accepts the context $\mathcal{C} = \{\mathbf{y}_1, \mathbf{s}_1, \ldots, \mathbf{y}_N, \mathbf{s}_N\}$ and received signal $\mathbf{y}$ as inputs, and produces an estimate $\hat{\mathbf{s}}$ of the transmitted symbol ${\mathbf{s}}$. To this end, first, the $N_r\times 1$ complex-valued received signal $\mathbf{y}$ is mapped to $2N_r\times 1$ real-valued vector $[\Re(\mathbf{y});\Im(\mathbf{y})]=\mathbf{y}^{\text{R}}$
by separating and concatenating real and imaginary parts, where $[\;\cdot\;;\;\cdot\;]$ denotes the operation of concatenating two column vectors. Each entry of vector $\mathbf{y}^{\text{R}}$ is normalized within the range $[0,1]$ as  $\overline{\mathbf{y}}=(\mathbf{y}^{\text{R}}-l_{\text{min}}\mathbf{1})/(l_{\text{max}}-l_{\text{min}})$,
where $\mathbf{1}$ is an all-ones vector, and we recall that $(l_{\text{max}},l_{\text{min}})$ is the dynamic range adopted by the quantizer $Q_b(\cdot)$ in \eqref{eq:rx}. This procedure is also applied to the received pilot signal $\{\mathbf{y}_i\}_{i=1}^N$ in the context. Vectors $\mathbf{y}$, $\{\mathbf{y}_i\}_{i=1}^N$, and $\{\mathbf{s}_i\}_{i=1}^N$ are zero-padded as needed to obtain $D_t=\max(N_t,2N_r)\times1$ vectors.
% The known pilot data $\mathbf{s}_i$ are normalized by 
% \begin{equation}
%     \overline{\mathbf{s}_i}=\frac{\mathbf{s}_i}{M}.
% \end{equation}
% Subsequently, the dimension of the shorter vector among $\overline{\mathbf{y}}$, $\overline{\mathbf{y}_i}$, and $\overline{\mathbf{s}_i}$ is extended to $D_t=\max(N_t,2N_r)$ through zero padding. 
The zero-padded sequences $\hat{\mathbf{y}}$, $\{\hat{\mathbf{y}}_i\}_{i=1}^N$, and $\{\mathbf{y}_i\}_{i=1}^N$ are then separately encoded into spiking signals via Bernoulli encoding \eqref{eq_bernoulli}, producing binary time sequences $\mathbf{y}^t$, $\{\mathbf{y}^t_i\}_{i=1}^N$, and $\{\mathbf{s}^t_i\}_{i=1}^N$, with $t=1,\cdots,T$, respectively.

\subsection{Spiking Token Embedding}
Each $D_t\times1$ vector in the set $\{\mathbf{y}^t_1,\mathbf{s}^t_1,\ldots,\mathbf{y}^t_N,\mathbf{s}^t_N,\mathbf{y}^t\}$ is considered a token. The spiking token embedding layer in Fig.~\ref{model} transforms each token into a vector of dimension $D_e\times1$ using a LIF layer with a $D_e \times D_t$ trainable embedding matrix $\mathbf{W}_e$. This produces ${2N+1}$ embeddings as
\begin{equation}
    \mathbf{E}_0^t = \text{LIF}^t([\mathbf{W}_e\mathbf{y}_1^t;\; \mathbf{W}_e\mathbf{s}_1^t;\; \ldots\; \mathbf{W}_e\mathbf{y}_N^t;\; \mathbf{W}_e\mathbf{s}_N^t;\; \mathbf{W}_e\mathbf{y}^t])
\end{equation}
The resulting post-embedding sequence at time step $t$, $\mathbf{E}_0^t$, comprises $M=2N+1$ columns, each corresponding to an embedded token with $D_e$ binary entries.

\subsection{Transformer Decoder Layers}
The spike embedding $\mathbf{E}_0^t$ undergoes processing through $L$ successive transformer decoder layers. Each $l$-th layer, with $l=1,\ldots,L$, accepts the $D_e\times M$ binary output $\mathbf{E}_{l-1}^t$ from the previous layer as input, and outputs a $D_e\times M$ binary matrix $\mathbf{E}_l^t$. The first sub-layer implements a masked multi-head spiking self-attention mechanism, whereas the subsequent sub-layer consists of a two-layer fully connected feed-forward network. Residual connections are utilized around each sub-layer, followed by layer normalization and a LIF neuron model.

\subsubsection{Masked Multi-Head Spiking Self-Attention Mechanism}
The first sub-layer deploys $n_h$ attention ``heads''. For the $h$-th head, the input $\mathbf{E}^t_{l-1}$ is transformed into query $\mathbf{Q}_h^t$, key $\mathbf{K}_h^t$, and value $\mathbf{V}_h^t$ matrices by applying separate encoding LIF layers as $\mathbf{Q}_h^t = \text{LIF}^t(\mathbf{W}_Q\mathbf{E}^t_{l-1})$, $\mathbf{K}_h^t = \text{LIF}^t(\mathbf{W}_K\mathbf{E}^t_{l-1})$, and $\mathbf{V}_h^t = \text{LIF}^t(\mathbf{W}_V\mathbf{E}^t_{l-1})$,
where $\mathbf{W}_Q$, $\mathbf{W}_K$, and $\mathbf{W}_V$ represent trainable    $D_K\times D_e$ real-valued    weight matrices, with $D_K = D_e / n_h$ assumed to be an integer. Subsequently, dot-product attention is executed through a specially designed architecture termed masked stochastic spiking attention (MSSA) \cite{song2024stochastic}, as detailed in Algorithm \ref{al1}.

\begin{algorithm}
\caption{Masked Stochastic Spiking Attention (MSSA)}
\begin{algorithmic}[1]
\State \textbf{Input:} $D_K\times M$ binary matrix sequences $\mathbf{Q}^t, \mathbf{K}^t, \mathbf{V}^t$ for $t=1,\ldots,T$, dimension $D_K$ (corresponding to one head $h$)
\State \textbf{Output:} $D_K\times M$ binary matrix sequence $\mathbf{F}^t$ for $t=1,\ldots,T$ (corresponding to one head $h$)
\For{$t = 1$ to $T$}
    \For{each $(m,m')$ in $M \times M$}
        \State Compute the query-key dot product for the $(m,m')$-th entry of the attention matrix:
        \quad\If{$m\leq m'$}
        \State$\tilde{A}^t_{m,m'} = \sum_{d=1}^{D_K} Q^t_{d,m} \land K^t_{d,m'}$
        \Else 
        \State $\tilde{A}^t_{m,m'} = 0$ (casual mask)
        \EndIf
        \State Normalize and encode $\tilde{A}^t_{m,m'}$ to obtain the binary attention parameter $A^t_{m,m'}$:
        \State\quad $A^t_{m, m'} \sim \mathrm{Bern}\left(\frac{1}{D_K} \tilde{A}^t_{m,m'}\right)$
    \EndFor
    \For{each $(d,m)$ in $D_K\times M $}
        \State Compute the $d$-th entry of each $m$-th output token as
        \State\quad $\tilde{F}_{d,m} = \sum_{m'=1}^{M} A^t_{m, m'} \land V^t_{d,m'}$
        \State Normalize and encode $\tilde{F}_{d,m}$ to obtain $F^t_{d,m}$ as
        \State\quad $F^t_{d,m} \sim \mathrm{Bern}\left(\frac{1}{M} \tilde{F}_{d,m}\right)$
    \EndFor
\EndFor
\end{algorithmic}
\label{al1}
\end{algorithm}

The deployment of MSSA leverages the binary nature of the inputs, enabling the efficient realization of the dot-product attention mechanism using simple logic AND gates and counters in lines 7 and 16 in Algorithm \ref{al1}. This circumvents the need for computationally expensive multiplication units. A potential hardware implementation of MSSA can be achieved by modifying the approach described in \cite{song2024stochastic}.

By Algorithm \ref{al1}, the output of the first sub-layer is obtained by the column-wise concatenation of the outputs from each head $h$, obtaining the binary $D_e\times M$ matrix
\begin{equation}
    \mathbf{G}^t = \left[\mathbf{F}_1^t \;;\; \mathbf{F}_2^t \;;\; \ldots \;;\; \mathbf{F}_{n_h}^t\right].
\end{equation}

\subsubsection{Feed-Forward Network}
The feed-forward network applies two sequential LIF layers, as described by
\begin{equation}
    \mathbf{E}^t_l = \text{LIF}^t(\mathbf{W}_2\text{LIF}^t(\mathbf{W}_1\mathbf{G}^t)),
\end{equation}
where the trainable real-valued matrices $\mathbf{W}_1$ and $\mathbf{W}_2$ are of dimensions $D_h \times D_e$ and $D_e \times D_h$, respectively. The output $\mathbf{E}^t_l$ constitutes the output of the $l$-th transformer decoder layer.

\subsection{Output Layer and Accumulation}
The output layer consists of a linear transformation that projects the embedding dimension into the number of classes $K^{N_t}$ via a trainable real-valued $K^{N_t}\times D_e$ matrix $\mathbf{W}_O$ as $\mathbf{O}^t = \mathbf{W}_O\mathbf{E}^t_L$. The outputs across time steps are accumulated and averaged to produce the final $K^{N_t} \times (2N+1)$ output $\mathbf{O}$:
\begin{equation}
    \mathbf{O} = \frac{1}{T} \sum_{t=1}^{T} \mathbf{O}^t = [\mathbf{o}_1 \; \mathbf{o}_2 \; \ldots \; \mathbf{o}_{2N+1}].
\end{equation}
The network's output is represented by the last token, $\hat{\mathbf{s}}=\mathbf{o}_{2N+1}$, with the index of the largest element being the symbol prediction.

\subsection{Off-Line Pre-Training}
For off-line pre-training, we assume the availability of a dataset with $N_{\text{Train}}$ pairs of fading channel coefficients and noise variance, $\{(\mathbf{H}_n,\sigma^2_n)\}_{n=1}^{N_{\text{Train}}}$, which are sampled i.i.d.. For each pre-training task, we are provided $N_{\text{Example}}$ pairs of context $\mathcal{C}$ and test input-output $\{\mathbf{y},\mathbf{s}\}$. The objective of pre-training is to minimize the cross-entropy loss averaged over all training tasks with respect to the model parameters $\theta$. The training loss function is accordingly defined as
\begin{equation}
L(\theta)=\sum_{n=1}^{N_{\text{Train}}}\sum_{l=1}^{N_{\text{Example}}}\left[-\sum_{c=1}^{K^{N_t}} f(\mathbf{s}_n)[c] \log\left(\hat{\mathbf{s}}[c]\right)\right],
\end{equation}
where $f(\cdot)$ denotes the one-hot encoding function that converts $\mathbf{s}_n$ into a binary one-hot vector of length $K^{N_t}$, and the notation $[c]$ signifies the selection of the $c$-th element.

% During the off-line phase 
% ICL represents a capability wherein a pre-trained sequence model, e.g. transformers, applies learned knowledge to new tasks without task-specific adaptation. When contextualized within MIMO systems, ICL facilitates the rapid interpretation of complex channel conditions. This is achieved by presenting the model with a context $\mathcal{C}$ composed of pilot signal-symbol pairs \( (y_i, s_i) \) that encapsulate latent channel information, allowing the model \( f_{\theta} \) to produce precise predictions \( \hat{s} = f_{\theta}(y | C) \) upon receiving a new data signal $y$. This approach obviates the need for iterative fine-tuning on pilot data, thus conserving computational resources and enabling rapid response. 

\section{Experimental Results and Conclusions} \label{se_eval}

This section evaluates the effectiveness of our SNN-based transformer architecture for MIMO symbol detection. We compare the performance and energy efficiency of the SNN-based implementation with an established benchmark algorithm, as well as a baseline ANN-based implementation of the ICL detector proposed in \cite{zecchin2023context}.
\subsection{Experimental Setup}

We investigate a MIMO system configuration featuring $N_t = N_r = 2$ antennas. We set $K=4$, so that the channel input utilizes a QPSK modulation scheme. A mid-tread uniform 4-bit quantizer, with clipping boundaries at $l_{\text{min}} = -4$ and $l_{\text{max}} = 4$, quantizes the received signal as \eqref{eq:rx}. The channel matrix elements adhere to an i.i.d. complex Gaussian distribution $\mathcal{CN}(0,1)$, and the signal-to-noise ratio (SNR) is uniformly sampled in the range of $[0, 30]$ dB.
  
The proposed symbol detection model utilizes an SNN transformer characterized by an embedding dimension $D_e = 256$, a total of $L = 4$ decoder layers, and $n_h = 8$ attention heads. The model undergoes training on a pre-training dataset comprising of $N_{\text{Train}} = 2^{15} = 32768$ combinations of tasks $(\mathbf{H}_n, \sigma^2_n)$. The context $\mathcal{C}$ includes $N = 20$ examples.

% \begin{figure*}[t]
%   \centering
%     \hspace{-5mm}\subfigure[]{\includegraphics[width=0.33\textwidth]{figures/results_ntasks_acc.png}\label{fig:1}}\hspace{-1mm}
%     \subfigure[]{\includegraphics[width=0.33\textwidth]{figures/results_computing_acc.png}
%     \label{fig:2}}\hspace{-1mm}
%     \subfigure[]{\includegraphics[width=0.33\textwidth]{figures/results_memaccess_acc.png}
%     \label{fig:3}}
%   \caption{(a) Detection BER for ANN and SNN transformers, pre-trained with varying numbers of tasks and tested under an SNR of 10 dB. Both models use $L=4$ layers, $n_h=8$ heads, and an embedding dimension of $D_e=256$. For the SNN model, the number of timesteps is set to $T=4$. (b) Computing energy consumption and (c) memory access energy consumption for ANN and SNN transformers of varying sizes, denoted by $L$-$D_e$, compared against the lowest BER achievable with a pre-training size of $N_{\text{Train}} = 32,768$. For the SNN model, $T=4$.}
%   \vspace{-5mm}
% \end{figure*}

\begin{figure*}[t]
  \centering
    \subfigure[]{\includegraphics[width=0.39\textwidth]{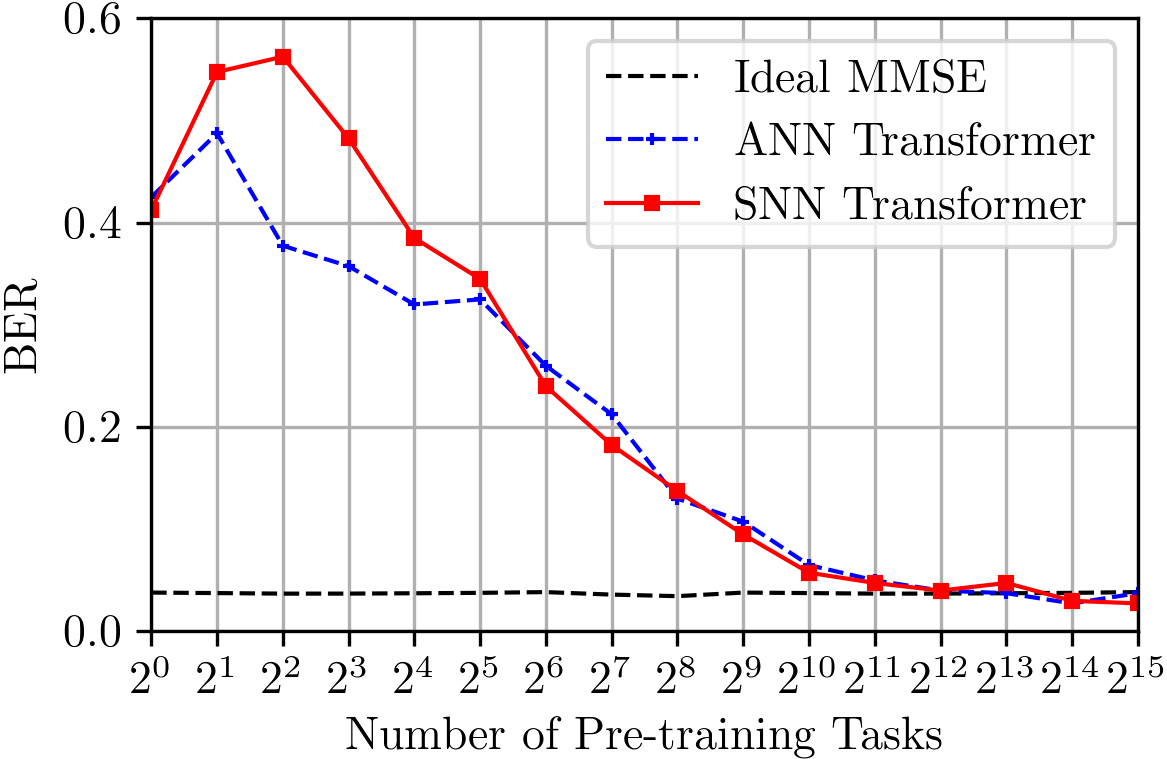}\label{fig:1}}\hspace{-1mm}
    \hfill\subfigure[]{\includegraphics[width=0.55\textwidth]{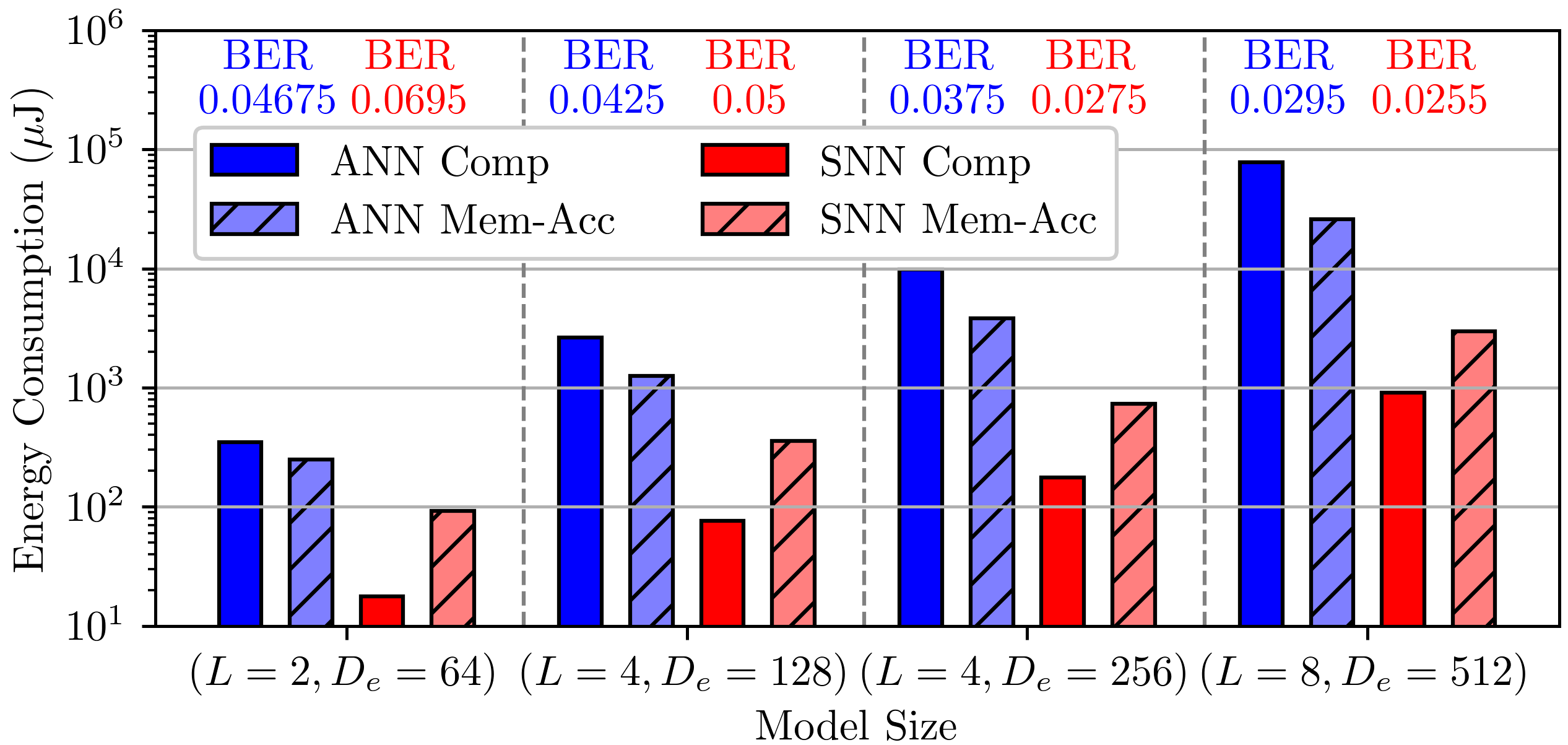}
    \label{fig:2}}\hspace{-1mm}
    \vspace{-3mm}
  \caption{(a) Detection BER for ANN and SNN transformers, pre-trained with varying numbers of tasks and tested under an SNR of 10 dB. Both models use $L=4$ layers, $n_h=8$ heads, and an embedding dimension of $D_e=256$. (b) Computing energy consumption and memory access energy consumption for ANN and SNN transformers of varying sizes, marked with the lowest BER achievable with a pre-training size of $N_{\text{Train}} = 32768$. For the SNN model, the number of timesteps is set to $T=4$ for both panels. }
  \vspace{-5mm}
\end{figure*}

The baselines include \textit{1) Ideal MMSE}: The MMSE equalizer with knowledge distribution of the testing channel $\mathbf{H}$ and noise variance $\sigma^2$; and \textit{2) ANN transformer}: the ANN counterpart of the proposed SNN transformer which follows the architecture as \cite{zecchin2023context} and shares the same size parameters as the SNN.

\subsection{Performance Test of ICL-based Symbol Detection} We first study the generalization performance of the ICL-based symbol detector as a function of the number of pre-training tasks $N_{\text{Train}}$.
Fig.~\ref{fig:1} presents the bit error rate (BER) for our SNN-based symbol detector as compared to its ANN-based counterpart. The BER is evaluated across a range of pre-training tasks, exponentially increasing from \(2^0\) to \(2^{15}\). The SNN transformer demonstrates performance comparable with the ANN transformer when the number of training tasks is sufficiently large, i.e., when $N_{\text{Train}}\geq 2^{5}$. Both implementations gradually approach the performance of the ideal MMSE equalizer. %\hl{explain BER vs network size tradeoff here for both ANN and SNN. for same model size, it appears SNN (4-256) has better accuracy than ANN.}

\subsection{Energy Efficiency}

We now estimate the energy consumption associated with the execution of ICL inference for the SNN-based symbol detector and its ANN-based counterpart. This estimation accounts for all necessary computational and memory access operations, both read and write, as per the methodology presented in \cite{acesnn2022}. In addition to the network size $(L,D_e)=(4,256)$, we examine the configurations $(2,64)$, $(4,128)$, and $(8,512)$ to evaluate the tradeoffs between performance and energy usage. The number of SNN time steps is set to $T=4$. For both ANNs and SNNs, model parameters and pre-activations are quantized using an 8-bit integer (INT8) format. The energy consumption calculations are based on the standard energy metrics for 45 nm CMOS technology reported in \cite{pedram2017dark,buffa2021voltage}.
We assume that the required data for computations are stored in the on-chip static random-access memory (SRAM), with a cache that is sufficiently sized to hold all model parameters, thus allowing for each parameter to be read a single time during the inference process.

Fig.~\ref{fig:2} shows the energy consumption evaluated for ANN- and SNN-based implementations with different model sizes. The energy consumption is broken down into separate contributions for computational processes and for  memory access. Each model size and implementation is marked with the BER attained at an SNR of 10 dB. 
% Fig.~\ref{fig:2} and Fig.~\ref{fig:3} show the energy consumption for computational processes and memory access, respectively, as a function of the BER at an SNR of 10 dB for various sizes of the SNN-based symbol detector relative to its ANN counterpart. 
We observe that the SNN-based implementation of size ($L=2$, $D_e=64$), exhibits a $20\times$ improvement in computational energy cost and $2.6\times$ saving in memory access cost as compared to its ANN-based counterpart, while achieving a worse BER, namely 0.070 against 0.047 for the ANN-based implementation.
 When increasing the model size to ($L=8$, $D_e=512$), the SNN-based detector significantly outperforms the ANN-based implementation with an $86\times$ reduction in computation energy and an $8.7\times$ reduction in memory access energy as compared to its ANN-based counterpart, while also achieving slightly better BER, namely 0.026 against 0.030. 
 
 %Overall, the SNNs demonstrate a reduction in total power consumption ranging from $5.4\times$ to $26.8\times$ depending on model sizes when compared with ANNs. 

%The demonstrated performance and energy efficiency of the SNN ICL-based symbol detector render it a suitable candidate for deployment on power-limited edge devices, aligning with the operational demands of 6G technology.

% \section{Conclusion}
% This paper introduces a novel adaptive neuromorphic computing framework using SNNs integrated with ICL for MIMO systems, specifically targeting the challenges of power efficiency and computational effectiveness.
\bibliographystyle{IEEEtran} 
\bibliography{reference}

\end{document}